\documentclass[a4paper,10pt]{article}

\usepackage[pdftex]{color,graphicx,hyperref}
\usepackage{amsmath,amsfonts,amssymb}
\usepackage{fullpage}
\usepackage{setspace}
\usepackage[labelfont=bf,labelsep=period]{caption}
\usepackage{subcaption}
\usepackage{cite}
\usepackage[utf8]{inputenc}
\usepackage[affil-it]{authblk}
\usepackage{color}

\newcommand{\fref}[1]{Fig.~\ref{#1}}

\title{Bridging the gap between graphs and networks}
\author[1,2,3,*]{Gerardo I\~{n}iguez}
\author[1]{Federico Battiston}
\author[1]{Márton Karsai}
\affil[1]{\normalsize{Department of Network and Data Science, Central European University, H-1051 Budapest, Hungary}}
\affil[2]{\normalsize{Department of Computer Science, Aalto University School of Science, 00076 Aalto, Finland}}
\affil[3]{\normalsize{IIMAS, Universidad Nacional Auton\'{o}ma de M\'{e}xico, 01000 Ciudad de Mexico, Mexico}}
\affil[*]{\normalsize{Corresponding author email: iniguezg@ceu.edu}}
\date{}

\begin{document}

\maketitle

\abstract{\small
Network science has become a powerful tool to describe the structure and dynamics of real-world complex physical, biological, social, and technological systems. Largely built on empirical observations to tackle heterogeneous, temporal, and adaptive patterns of interactions, its intuitive and flexible nature has contributed to the popularity of the field. With pioneering work on the evolution of random graphs, graph theory is often cited as the mathematical foundation of network science. Despite this narrative, the two research communities are still largely disconnected. In this Commentary we discuss the need for further cross-pollination between fields -- bridging the gap between graphs and networks -- and how network science can benefit from such influence. A more mathematical network science may clarify the role of randomness in modeling, hint at underlying laws of behavior, and predict yet unobserved complex networked phenomena in nature.
}



\subsection*{The mathematics of a networked reality}



Behind the history of mankind's scientific progress there is a story of mathematical theory building. One where scientists not only uncover the behavior of empirical phenomena through experiments and data analysis, but develop theories that abstract such behavior mathematically~\cite{popper2002popper}. This process amounts to scientific understanding when mathematics is \textit{sufficiently isomorphic} to reality~\cite{korzybski1990alfred}; when theory includes the mechanisms deemed most important, and despite ignoring some features, still allows for logical inferences that accurately describe the phenomena of interest~\cite{wigner1960unreasonable}. A sufficient similarity in structure between mathematics and reality allows scientists to forecast events, control natural systems, and even predict behavior that has not yet been confirmed empirically~\cite{lin1988mathematics}.

Up until recent times, scientific efforts towards a mathematical representation of reality have mostly focused on elements of the `simplest' natural systems at both the smallest and largest scales: the physical, chemical, and biological entities comprising the microscopic world, as well as thermodynamic systems and the large-scale structure of the Universe. The success of this endeavor benefits from the availability of reproducible data and the possibility of partially disregarding the variability of a system's components and their interactions, while keeping the theory accurate. It is mostly in the last decades that scientists have explored the mathematics of phenomena where elements are {\it mesoscopic}: the brain and the physiology of living bodies, ecosystems, and the social, technological, and economic constructs of our modern world~\cite{ball2004critical}. These systems are inherently complex in the sense that the (varying and often adaptive) interactions between components are an essential feature of a sufficiently isomorphic mathematical theory. Therefore, any framework disregarding heterogeneity of elements or interactions will fail in accurately describing and predicting the behavior of complex interacting systems~\cite{vespignani2009predicting}.

The need to mathematically describe both components and interactions has promoted the rise of network science as an interdisciplinary effort to quantify the structure and dynamics of complex systems~\cite{barabasi2016network,newman2018networks}. Facilitated by the big data revolution, the concept of network (a set of elements or \textit{nodes} connected by their interactions, or \textit{edges}) has become a powerful tool to describe the myriad of empirical observations in physical, biological, and social phenomena. With inspiration from seminal studies in sociology and economics~\cite{wasserman1994social,borgatti2009network,jackson2010social,easley2010networks}, network science inherited its first concepts from graph theory, the branch of mathematics concerned with pairwise relations between objects~\cite{lovasz2012large}. Since then, however, graph theory and network science have taken separate directions, with little overlap in their research questions and academic communities. Graph theory has focused on providing rigorous proofs for graph properties, such as graph enumeration, coloring and covering (with applications ranging from chemistry to circuit design)~\cite{bondy1976graph}. Current network science, instead, is more akin to phenomenological physics by focusing on observations of real-world networks and ad hoc mathematical concepts to quantify them, with the goal of gaining intuition of their underlying generative mechanisms. Due to its aim of pursuing rigorous arguments, graph theory has so far concentrated on structures that are more analytically treatable, like random or dense graphs, whereas network science focuses on the most common features seen in data, such as sparsity and inhomogeneities in the structure and temporal behavior of large but finite networks.

Rather than an unavoidable distance, we believe this gap is an unprecedented opportunity to bridge the two scientific disciplines together: a way of bringing empirically motivated questions to the attention of graph theory, and highlighting unknown results from graph theory to network science. Through wide-ranging, interdisciplinary projects like the ERC Synergy Grant `DYNASNET'~\cite{european2019dynamics}, scientists aim at connecting concepts of interest in graph theory (such as representative sets and limit objects~\cite{lovasz2006limits}) to the `big ideas' of network science, including heterogeneity, structural and dynamic correlations, communities and mesoscopic order, and the coupling between the dynamics of and on networks~\cite{fortunato2010community,holme2012temporal}. Echoing the way some fields of physics have become more mathematical over time, this effort will bring formality and precision to the mathematical constructs needed to quantify empirical results in network science. At the same time, graph theorists will be presented with new theoretical challenges brought by network scientists in the analysis of network control~\cite{liu2011controllability}, physical networks~\cite{dehmamy2018structural}, or networks with higher-order interactions~\cite{lambiotte2019networks}. We believe this synergy will enhance the ability of both disciplines to give quantitative predictions about the complex networked systems of nature and society.


\subsection*{Learning from mathematical physics}


Building a mathematical framework to uncover logical consistency in an empirical science is a well-known challenge in the history of physics, from the first theories of celestial and Newtonian mechanics, to electromagnetism, relativity, and quantum field theory. First, there are generic observations backed up by reproducible data and experiments, followed by attempts at mathematical description usually motivated by inconsistencies in previously accepted theories~\cite{popper2002popper,kuhn2012structure}. A feedback loop between more accurate observations and more sophisticated mathematics typically follows, ultimately leading to mathematical theories with predictive power.

In electromagnetism, the experimental prowess of Faraday and others eventually led to the rigor of Maxwell's equations, a mathematical framework that, in turn, predicted the invariance of the speed of light and served as intuition for Einstein's theory of relativity. In statistical mechanics, Wilson developed a renormalization group approach to derive the observed coincidence of critical exponents (within a universality class) near continuous phase transitions, explaining why some features of thermodynamic systems are insensitive to microscopic details. Currently, the mathematical theories of partial and stochastic differential equations are widespread in physics, with applications in electromagnetism, statistical mechanics, hydrodynamics, acoustics, elasticity theory, and astrophysics. An accurate mathematical theory can even indicate where (not) to look for new physics: Stemming from the analysis of constants of motion in Lagrangian and Hamiltonian mechanics, Noether's theorem (relating symmetries and conserved quantities in a system) is a powerful tool to find transformations that make physical laws invariant in time and space.

Network science and the developments in graph theory most related to empirical findings already feature attempts where rigorous mathematical frameworks, initially motivated by real-world observations, have led to valuable insight on the similarities between the underlying mechanisms of complex networked systems. The analysis of random graphs~\cite{erdHos1960evolution} showed that the slow growth of average path length with network size, colloquially know as the `small-world' effect, is not only ubiquitous in most real-world networks, but appears even in the simplest case of Bernoulli distributed edges. In turn, the exploration of large-scale social, biological and technological networks has prompted scientists to create alternatives to this formalism that account for empirically relevant features like degree heterogeneity~\cite{albert2002statistical,bollobas2007phase}, high clustering~\cite{watts1998collective} and community structure~\cite{fortunato2010community,karrer2011stochastic}. Inspired by the connection between spacetime geometry and matter distribution in general relativity, network scientists have also explored the relationship between embedding a network in a hyperbolic space and the appearance of structural correlations and heterogeneity~\cite{krioukov2010hyperbolic}. Finding connections between geometry and network structure may turn out to have immediate practical applications, such as enhanced network navigability thanks to new notions of distance in the underlying space.


\subsection*{The future breakthroughs of mathematical network science}


Motivated by the numerous results of large network analysis made recently, in the following we sketch a list of future breakthroughs that may aid in bridging the gap between the formal statements of graph theory and the data-driven intuition of network science. We speculate these breakthroughs will lead to the development of a more mathematical network science that, just like standard physical theories, will provide the mathematical foundation for a comprehensive understanding of complex networked systems.

\begin{figure}[t]
\centering
\includegraphics[width=0.7\textwidth]{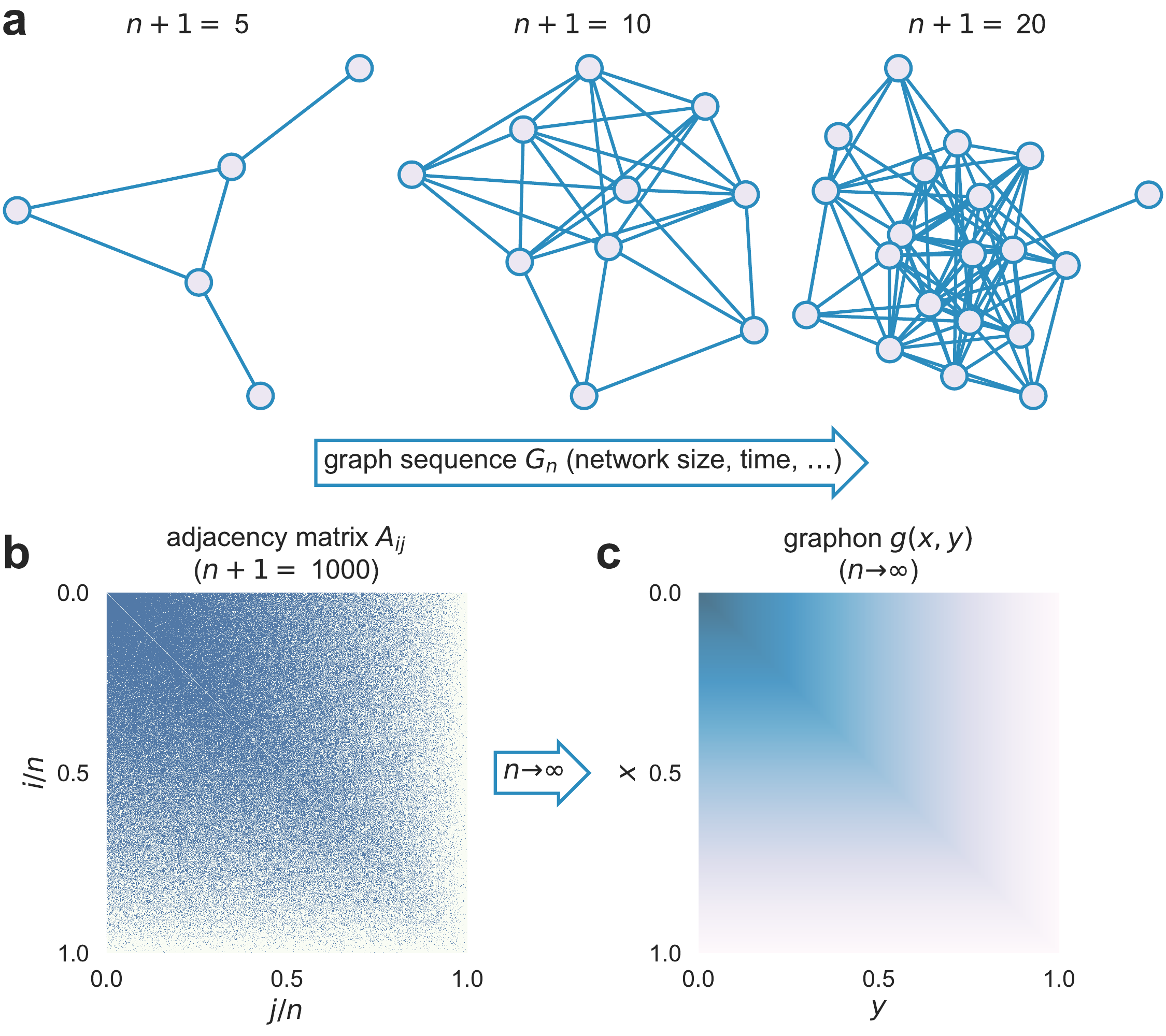}
\caption{
\small {\bf Limit objects of dense growing networks.} {\textbf{(a)}} Snapshots of a growing uniform attachment graph $G_n$ of size $n+1$. Starting from a single node at $n=0$, in each iteration $n$ a node is added and all pairs of non-connected nodes are linked with probability $1/n$. This is a simplified preferential attachment mechanism for dense networks reminiscent of those arguably driving the growth of empirical scale-free networks~\cite{albert2002statistical}. {\textbf{(b)}} Adjacency matrix $A_{ij}$ for large $n$ ($i, j = 0, \ldots, n-1$), showing how older nodes ($i \ll n$) are more connected than newer nodes. {\textbf{(c)}} In the limit $n \to \infty$, $A_{ij}$ tends to the \textit{graphon} $g(x, y) = 1 - \text{max}(x, y)$ for $i = xn$ and $j = yn$~\cite{lovasz2012large}. Many problems related to network structure can be stated and solved more easily by considering limit objects instead of finite-size networks.}
\label{fig:1}
\end{figure}

\paragraph{Limit objects for sparse graphs.} There is a growing literature on the properties of dense graph sequences and their associated limit objects, \textit{graphons}~\cite{lovasz2006limits} (see \fref{fig:1}). Graphons are useful in simplifying the formulation and proof of theorems related to graph distances, subgraph sampling and convergence of graph sequences~\cite{borgs2008convergent,borgs2012convergent}. However, most real-world networks are not dense but sparse, with relatively few edges between nodes. Once graph theory can describe the empirically relevant, asymptotic behavior of sparse graph sequences, these results will find applications in network science and complex systems forecasting. Promising limit objects include \textit{graphings} (for degree bounded graphs)~\cite{lovasz2012large} and \textit{graphexes} (for graphs as random measures)~\cite{caron2017sparse,borgs2019identifiability}.

\paragraph{Sparse graph theory of structural/temporal heterogeneity.} The study of sparse limit objects might also consider other structural and dynamical constraints widely documented in network science. Real-world networks display heterogeneous structure at several scales, with mesoscopic patterns of core-peripheries and communities~\cite{fortunato2010community}. Empirical networks are also temporal in the sense that nodes and edges can appear and disappear following a dynamics that is not homogenous~\cite{holme2012temporal}, and not necessarily deterministic. We expect graph theoretical advances such as limit objects for stochastic-block-model networks~\cite{karrer2011stochastic}, the analysis of graph sequences where the sequence index is time instead of graph size, and random matrix theory~\cite{mehta2004random} for sparse graphs.

\paragraph{Insights about dynamical processes on/of networks.} A large part of network science deals with the effect of network structure on dynamical processes such as epidemic spreading, information transfer, and synchronization. A related topic is the dynamics of the network itself: network growth and decay, edge rewiring, and temporal and adaptive networks~\cite{gross2008adaptive,holme2012temporal}. Current approaches include network-aware dynamical systems theory, linear stability analysis, and matrix spectral theory~\cite{porter2016dynamical}. Still, mathematical ideas concerning empirically relevant networked dynamics are needed, particularly when the dynamics on the network and of the network have similar time scales, or when networks have multiple layers or higher-order interactions~\cite{porter2019nonlinearity}. We believe advances in algebraic geometry, computational topology, and graph limit theory will prove useful in overcoming this challenge.


\subsection*{Bridging the gap between graphs and networks}


How can scientists achieve these theoretical breakthroughs, thus bridging the gap between graph theory and network science? We once more turn to the history of mathematical physics for clues. In the early stages of thermodynamics, the likes of Carnot, Clausius, and Kelvin discovered phenomenological relations between measurable quantities of large physical systems, without yet a full understanding of the underlying dynamics. Then, giants like Boltzmann, Maxwell, and Gibbs used probability theory to derive these relations from microscopic laws of mechanics, implying that macroscopic observations are the statistical outcome of more fundamental processes at lower scales. With its many observations and simplified phenomenological models, we see current network science in a state similar to 19th century thermodynamics. A stronger link between graph theory and network science will allow researchers to refine idealized assumptions on networked phenomena, and find mathematical connections between these microscopic mechanisms and large-scale behavior with even more relevance for real-world applications.

In the next years, members of `DYNASNET'~\cite{european2019dynamics} (and similar upcoming projects) will work on various approaches to bridge the efforts of graph theory and network science. One possibility is to use graph theory results as selection principles for microscopic laws of behavior in network science, just like Noether's theorem identifies classes of hypothetical Lagrangians that could describe an empirically invariant quantity. These principles would not only make modeling efforts more efficient; they would highlight apparently unconnected phenomena driven by the same underlying mechanisms. Conversely, limit objects in graph theory might lead to network versions of the central limit theorem: the mathematical derivation of features shared by many large-scale complex systems despite their differing elements and interactions. Another possibility is to further explore the role of stochasticity in complex networked systems. Network scientists use randomness as a substitute for unknown rules of interaction, and then infer macroscopic behavior mostly via numerical and data-driven simulations. Empirically adjusted versions of graph theoretical results, such as Szemerédi's regularity lemma~\cite{lovasz2007szemeredi}, might uncover the fundamental interplay between randomness, structure, and dynamics in real-world networks.

In the future, researchers and scientifically minded policy makers may use these advances to make quantitative predictions of yet unobserved complex networked phenomena.  Network control theory has already found previously unknown neurons related to locomotive behavior in the \textit{C. Elegans} connectome~\cite{yan2017network}. Network epidemiology is steadily approaching the ability to deliver operational forecasts of worldwide epidemic outbreaks~\cite{viboud2019future}, and to predict the effect of emergency policies such as travel restrictions~\cite{chinazzi2020effect}. Some of the future predictions stemming from the engagement of graph theory and network science may even follow the steps of giant theoretical achievements such as the discovery of the Higgs boson in the Standard Model, or the existence of gravitational waves in general relativity. If networks are indeed a fundamental facet of physical reality at mesoscopic scales, as many of us believe, undiscovered consequences of their presence should be everywhere. We anticipate these future discoveries will connect seemingly distant disciplines in an overarching network of scientific knowledge tempered by mathematics.


\subsection*{Acknowledgments}

{\small
We thank M. Abert, J. Kert{\'e}sz, T. Peixoto, and M. P{\'o}sfai for helpful discussions and comments during the writing of this manuscript. We acknowledge support from DYNASNET ERC Synergy Grant (ID: 810115).
}

\bibliographystyle{ieeetr}
\bibliography{references}

\end{document}